# PROGRESS IN INDUCTION LINACS

George J. Caporaso
Lawrence Livermore National Laboratory, Livermore, California 94550 USA

*Abstract*

This presentation will be a broad survey of progress in induction technology over the past four years. Much work has been done on accelerators for hydrodynamic test radiography and other applications. Solid-state pulsers have been developed which can provide unprecedented flexibility and precision in pulse format and accelerating voltage for both ion and electron induction machines. Induction linacs can now be built which can operate with MHz repetition rates. Solid-state technology has also made possible the development of fast kickers for precision control of high current beams. New insulator technology has been developed which will improve conventional induction linacs in addition to enabling a new class of high gradient induction linacs.

## 1 INTRODUCTION

The last several years have seen dramatic advances in linear induction accelerator technology. There have been revolutionary advances in pulsed power drivers for both accelerators and fast kickers that now offer unprecedented speed, pulse format flexibility and voltage precision. Some of this new technology will be used in the $2^{nd}$ axis of DARHT. New compact focusing lenses have been developed for Heavy Ion Fusion drivers and advances in insulators and dielectrics for pulse forming lines offer the possibility to realize a new class of induction accelerators with gradients an order of magnitude higher than the current state of the art.

## 2 ADVANCED RADIOGRAPHY AS A DRIVER OF INDUCTION TECHNOLOGY

A significant impetus for the development of advanced technology has come from the area of flash x-ray radiography for stockpile stewardship. The cessation of underground nuclear testing has placed increased emphasis on flash x-ray radiography. In order to meet the need for acquiring multiple line-of-sight, multiple time frame data during a single hydrodynamic test a concept was developed that provides these features *using a single accelerator* [1].

By equipping a long pulse induction linac with a pulsed power source capable of running in the MHz range and using fast kickers multiple pulses and many lines of sight could be achieved.

This concept is illustrated in Figure 1 which shows the key elements of advanced technology necessary to make the scheme viable. A solid-state modulator powered by Field Effect Transistors (FETs) is shown along with a fast kicker capable of switching kiloampere level beams into different transport lines.

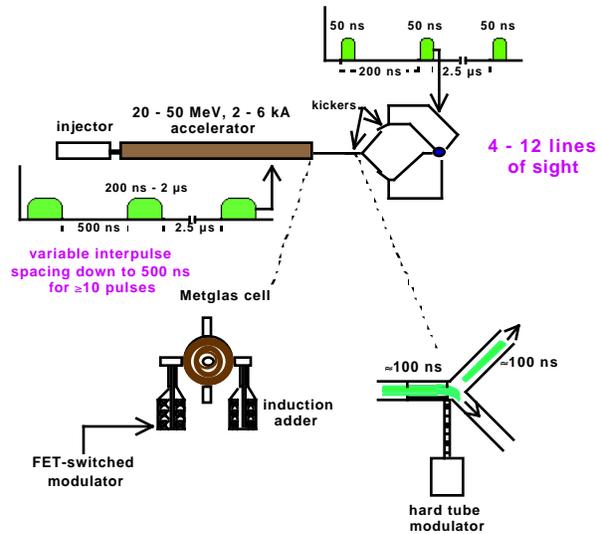

Fig. 1. Linear induction accelerator concept for multi-axis, multi-time frame flash x-ray radiography.

## 2. A. SOLID-STATE MODULATOR

The basic concept of the solid-state modulator is shown in figure 2. The accelerating voltage is established on a large capacitor bank which is connected (and disconnected) from the load by a fast switch, which is this case is a series-parallel array of FETs. The load in this case is an induction core.

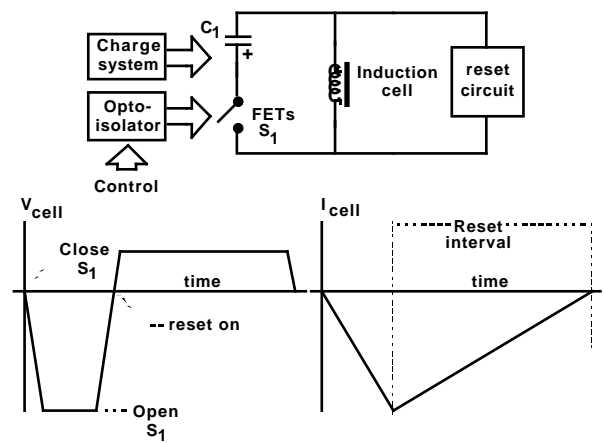

Fig. 2. Basic circuit concept for the solid-state modulator. A fast switch (FET arrays) connect and disconnect a pre-charged capacitor to the induction cell. Note that the output voltage is negative in this illustration.

The induction core permits these circuits to be stacked in an inductive voltage adder configuration. A fast reset circuit is also provided to enable the modulator to operate at very high repetition rates.

This circuit concept allows the pulse format to be changed without changing the circuit configuration. Simply by programming the optical trigger sequence the width and inter-pulse time for each pulse can be changed at will. A single layer of the ARM-II modulator is shown in Figure 3.

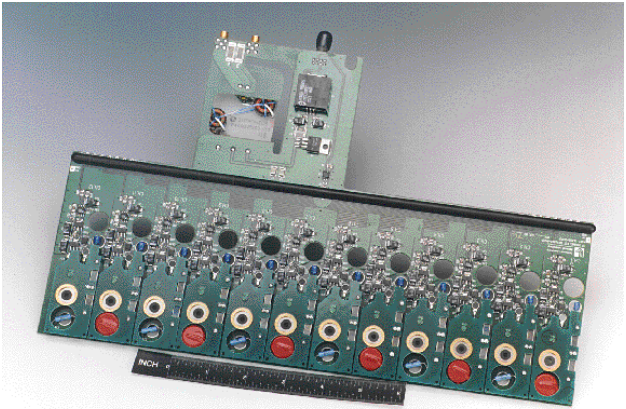

Fig. 3. A single switch board of the ARM-II modulator. The board contains 12 FETs which switch 100 Amps each at roughly 800 volts. These circuit boards are stacked in series to provide 15 kV for the ARM-II modulator. There are four of these stacks in parallel around a Metglas core to provide 4.8 kA.

The circuit boards shown in Figure 3. Are stacked in series and parallel to provide a modulator capable of running at 1 MHz with full reset and at 2 MHz without reset.

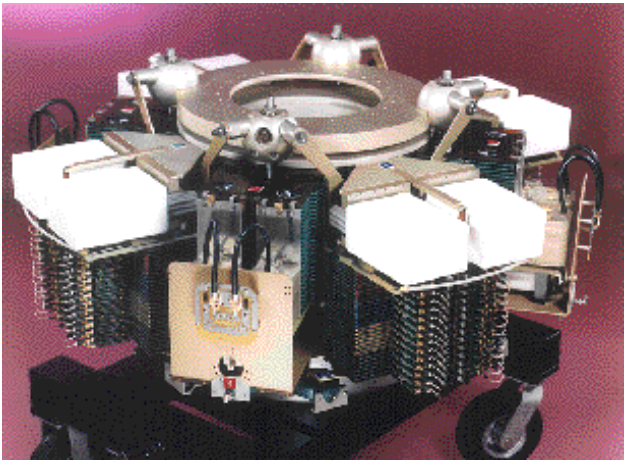

Fig. 4. A single ARM-II modulator capable of generating a programmable burst of variable width pulses at 15 kV (open circuit voltage) and 4.8 kA output current.

The ARM-II modulator was designed to be stacked in an inductive voltage adder configuration. A three-stage adder is shown in Figure 5.

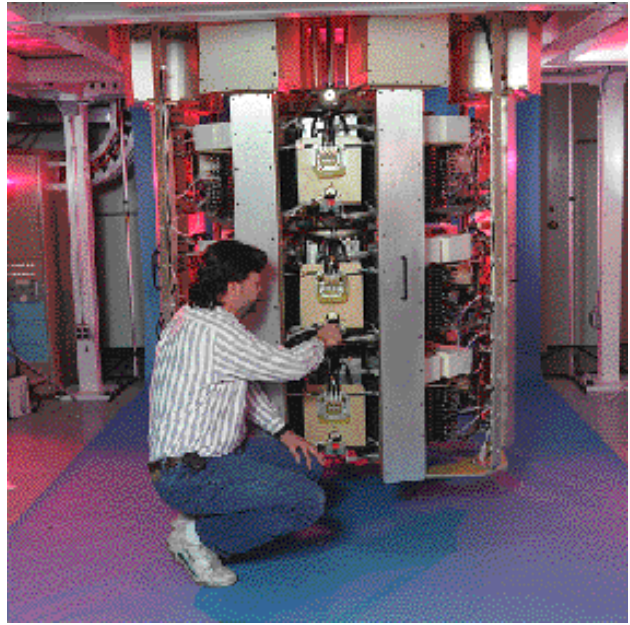

Fig. 5. The three-stage ARM modulator assembly which produces 45 kV (open circuit) at 4.8 kA and 1 MHz repetition rate.

A typical burst which illustrates the pulse format flexibility is shown in Figure 6.

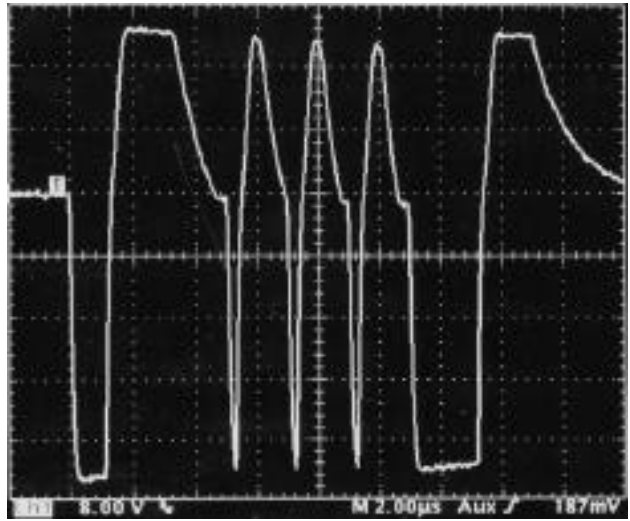

Fig. 6. Typical burst from the modulator showing the pulse format agility of the system. The horizontal scale is 2 μsec. per division.

This technology will also be used to power the high current kicker for the 2$^{nd}$ axis of DARHT.

## 2. B. FAST, HIGH-CURRENT KICKER

The second major piece of technology inspired by the needs of Advanced Radiography is a precision, fast kicker capable of handling long multi-kiloamp beam pulses. In order to obtain switching times of order 10 ns. the source of the fields must be inside the beam pipe. A configuration was chosen which is similar to a stripline configuration similar to that of a stripline beam position monitor was chosen. The concept is shown in Figure 7.

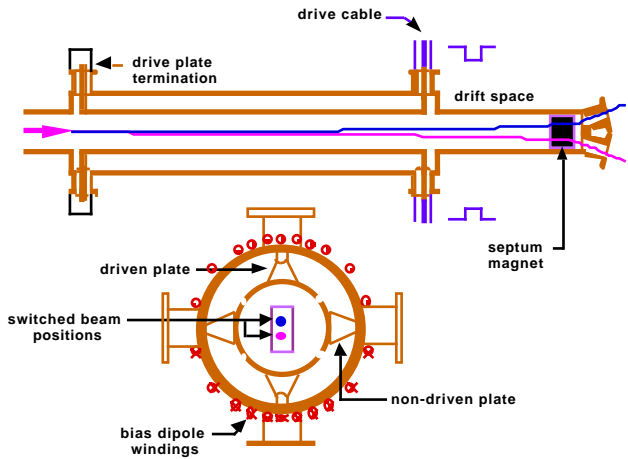

Fig. 7. Fast, high-current kicker concept. The system consists of four equal size striplines. One opposite pair of electrodes is powered to produce switching in a plane. In this illustration the top electrode is powered negatively while the bottom electrode is powered positively to switch the beam downward in the vertical plane. Since the pulser technology employed is unipolar, a D.C. bias dipole winding wrapped over the kicker vacuum housing pre-steers the beam upwards to obtain a full range of vertical motion.

The system that has undergone extensive testing on the ETA-II accelerator at Livermore is shown in Figure 8.

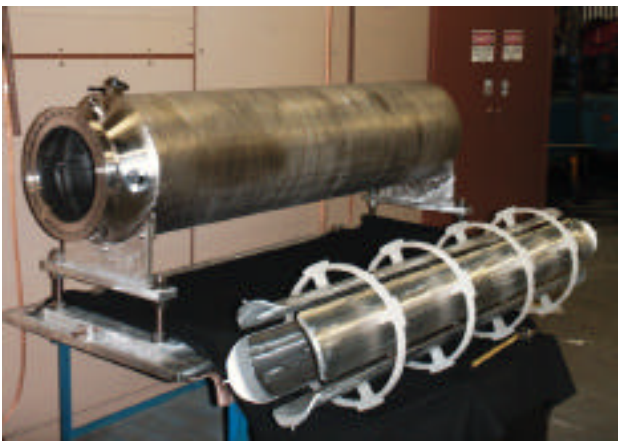

Fig. 8. Stripline kicker system used on ETA-II. A system very similar to this will be deployed on the second axis of DARHT to produce a sequence of radiographic pulses from a 2 μsec pulse in the accelerator.

The kicker system has been very successful. Beams of up to 2 kA, 50 ns wide at 6 MeV have been steered rapidly with precision. An image from the switched beam intercepting a quartz foil is shown in Figure 9. The picture captures a single ETA-II pulse in the act of being switched from one position to another.

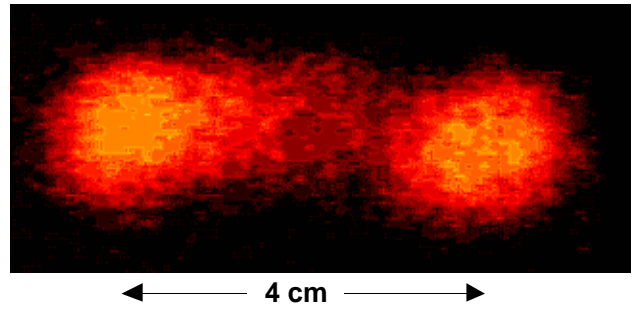

Fig. 9. A single beam pulse from ETA-II caught in the act of switching from an initial position on the right to the final position on the left. The image is created by light striking a quartz foil approximately 60 cm downstream of the end of the kicker. The total centroid shift is some 4 cm with a kicker plate voltage of ± 9 kV.

## 2. C. DARHT

The Dual Axis Radiographic Hydrodynamic Test (DARHT) facility is under construction at Los Alamos. It consists of a single 70 ns pulse, 20 MeV, 4 kA induction linac which is in operation and a second, long pulse machine under construction now. The second axis (DARHT-2) will produce a 2 μsec, 20 MeV pulse at up to 4 kA. A kicker system will be used to extract a sequence of 4 relatively short radiographic pulses out of the long pulse and direct these to the x-ray converter target.

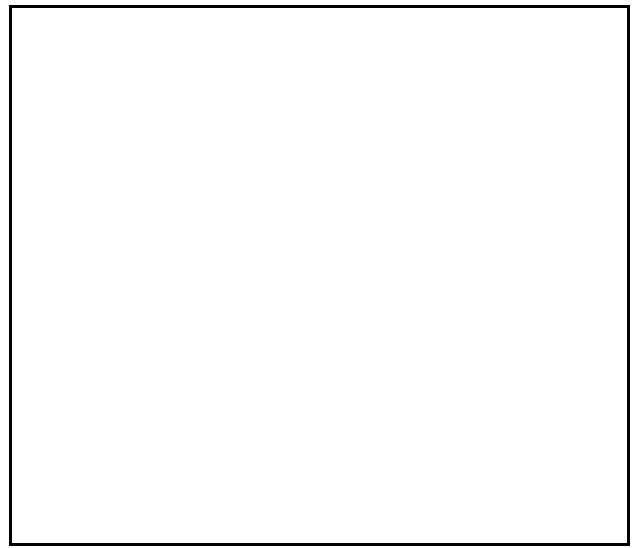

Fig. 10. The DARHT facility at LANL. The axes of the two radiographic machines are at right angles.

## 3 ADVANCES IN HEAVY ION FUSION TECHNOLOGY

There have been numerous advances in the technologies required for Heavy Ion Fusion driver development. Very compact superconducting lenses have been developed which will lead to increased accelerating gradients. An example of such a lens is shown in Figure 11.

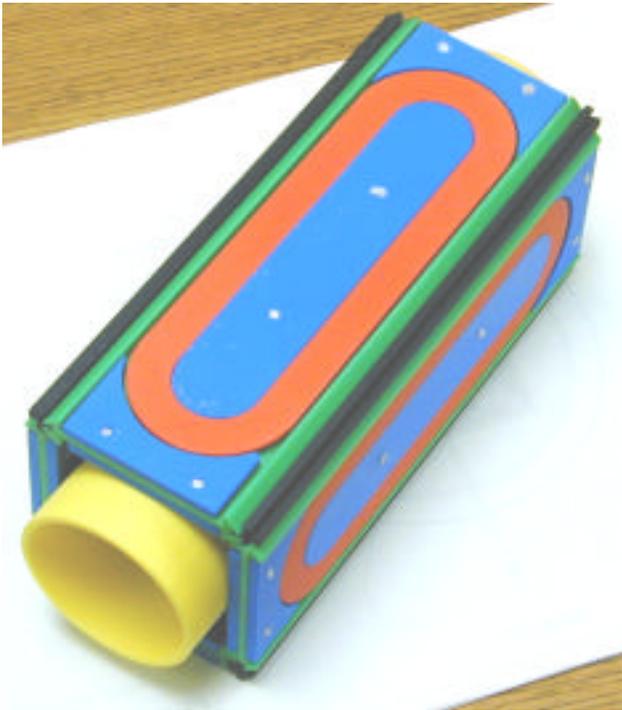

Fig. 11. Superconducting quadrupole using a Rutherford cable on a flat support (Martovetsky at LLNL).

Compact pulsed magnets suitable for focussing arrays of beamlets have also been developed. A prototype array for the Integrated Research Experiment (IRE) is shown in Figure 12.

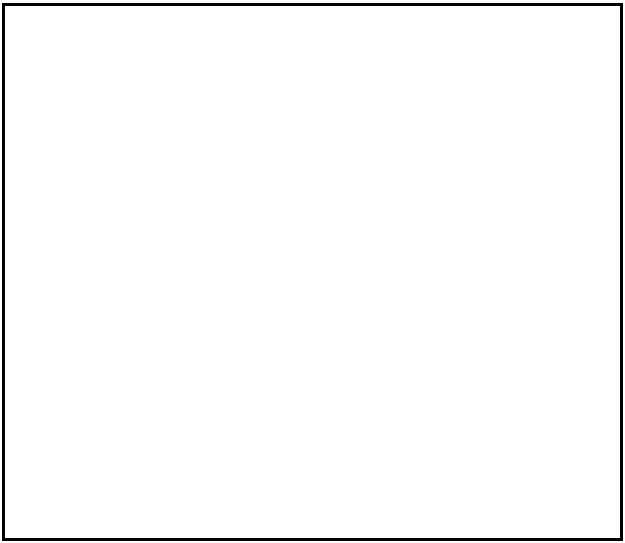

Fig. 12. Compact, pulsed quadrupole lens array for the IRE.

Improved characterization of magnetic core materials such as Metglas, Finemet and Silicon-Iron under conditions comparable to those found in a driver have been completed which will lead to the most efficient and economical choices for the several different accelerator cell systems employed in a driver.

## 4 ADVANCED INSULATORS AND THE DIELECTRIC WALL ACCELERATOR

## 4. A. HIGH GRADIENT INSULATORS

In the past few years a new class of insulators has been developed that has superior performance for short, long and bi-polar pulses. Called *simply high gradient insulator*s (HGI) these are novel configurations of conventional insulating materials [2].

The basic idea of the HGI is to interrupt the normal insulator with finely spaced, floating electrodes. The typical spacing between electrodes can be a few mm down to 0.1 mm. In general, the voltage holding ability of these configurations improves as the period length is shortened. Insulators have been fabricated from dielectrics such as kapton, rexolite, lexan and fused silica. A few samples are shown in Figure 12. These insulators have flashover strengths 2 to 5 times higher than conventional insulators.

A startling result is the excellent performance of these insulator configurations in the proximity of high current electron beams. One such test is shown in Figure 13.

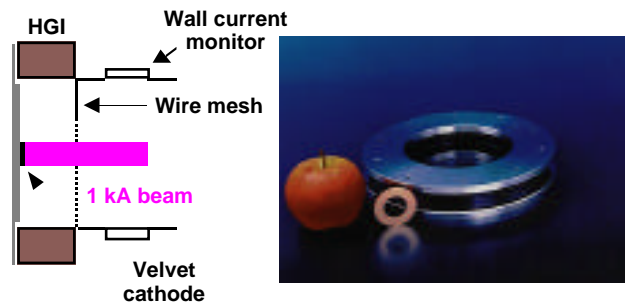

Fig. 13. High gradient insulator test using the kapton/stainless steel insulator shown on the right side of the figure. The insulator measured 22 cm outer diameter by 2 cm in axial length. A velvet cloth 1 cm in radius was used as a cathode while a highly transparent wire mesh was used as an anode. A 20 ns. FWHM, 440 kV pulse was placed across the outer diameter of the insulator. The cathode produced a 1 kA electron beam repeatedly with no breakdowns. As the voltage was increased signs of insulator breakdown at the end of the pulse could be observed. The breakdown-free accelerating gradient in the presence of this beam was 22 MV/Meter.

In another test of the insulator in the presence of beam an ETA-II induction cell was modified to accept a high gradient insulator. The standard insulator, a slanted piece of rexolite with a slant width of 3.75 cm was replaced with a high gradient version with a straight wall and only 1 cm wide.

This cell was installed on the end of ETA-II about 10 cm from a graphite beam stop. Voltage was applied to the cell by coupling the beam return current through load resistors. The cell took the full beam current (2 kA, 50 ns pulses at 6 MeV) at 1 Hz for an entire day. The cell logged over 10,000 shots with no breakdowns. The cell was operated at up to twice it's normal operating voltage with a straight wall insulator having a direct line of sight to the beam and a width almost one quarter of the standard insulator with no breakdowns at 17.5 MV/Meter.

These results suggest the possibility of an accelerator configuration that might have a considerably higher gradient than conventional induction accelerators.

## 4. B. THE DIELECTRIC WALL ACCELARATOR

The basic concept for a Dielectric Wall Accelerator (DWA) is shown in Figure 14. A conventional induction machine has an accelerating field only in the gap, which occupies a relatively small fraction of the axial length of an accelerating cell. If the conducting beam pipe could be replaced by an insulating wall, accelerating fields characteristic of the gaps might be applied uniformly over the entire length of the accelerator yielding a much higher gradient.

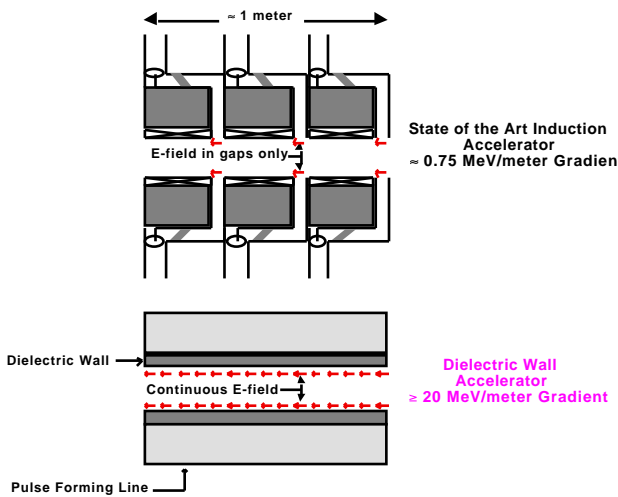

Fig. 14. Basic idea of the DWA.

In order to supply an accelerating voltage over the entire structure a suitable pulse forming line must be used along with a closing switch to initiate the voltage pulse. One such concept is shown in Figure 15.

The Asymmetric Blumlein is configured as two radial transmission lines with different dielectrics. These lines are initially charged to the same voltage but opposite polarities so that there is no net voltage across the pair of lines. If switches on the outer diameter are closed, waves will propagate radially inward leaving zero voltage in their wakes. Because the dielectrics in the lines have different values of permittivity, the waves travel at different speeds. When the faster of the two waves hits the inner boundary of the line there will be a reflection because of an impedance mismatch which will boost the voltage of that wave, causing the polarity of that line to reverse which generates a net accelerating voltage across the line I.D.

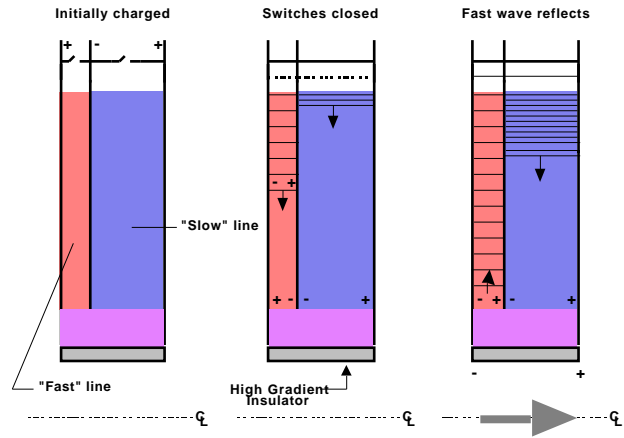

Fig. 15. The concept of the Asymmetric Blumlein one of several schemes for providing an accelerating pulse over the DWA.

## 5 CONCLUSIONS

Many important technological advances in the induction accelerator field have taken place over the past several years. A revolution in pulsed power technology has boosted the maximum repetition rate of induction machines by three orders of magnitude over the previous record providing unprecedented pulse format flexibility and voltage precision. Progress in compact magnetic lenses and lens arrays for Heavy Ion Fusion promise systems with higher average accelerating gradients. Advanced understanding of magnetic core material will permit the construction of more efficient and economical fusion drivers. A new class of high gradient insulators has demonstrated superior performance in a variety of modes and promises to make possible the construction of novel high gradient accelerators.

## 6 ACKNOWLEDGMENTS

It is a pleasure to acknowledge the help of many colleagues at LLNL, LANL, LBNL, Honeywell FM&T and elsewhere: Yu—Jiuan Chen, Judy Chen, Steve Sampayan, Jim Watson, John Weir, Ed Cook, Tim Houck, Glen Westenskow, Hugh Kirbie, Dave Sanders, Mike Burns, Mike Krough, John Barnard and Art Molvik.

This paper is dedicated to my friend Dan Birx who was a true genius and an outstanding human being who had a profound influence on this field.

This work was performed under the auspices of the U.S. Department of Energy by the University of California Lawrence Livermore National Laboratory under contract W-7405-ENG-48.